\documentclass[aps, prb, twocolumn,showpacs, final]{revtex4-1}
\usepackage{amsthm}
\usepackage{amsfonts}
\usepackage{amsmath}
\usepackage{amssymb}
\usepackage{mathtools}
\usepackage{alltt}
\usepackage{subfigure}
\usepackage{mathptmx}
\usepackage{fancyhdr}
\usepackage{amsbsy}
\usepackage{textcomp}
\usepackage{array}
\usepackage{fixmath}
\usepackage[english]{babel}
\usepackage{graphicx}
\usepackage{color}

\usepackage{epstopdf}

\begin{document}

\title{Order-disorder phase change in embedded Si nano-particles}
\author{Sergio Orlandini$^{1,2}$}
\author{Simone Meloni$^{2,3}$}\email{To whom correspondence should be addressed: s.meloni@caspur.it} 
\author{Luciano Colombo$^4$}
\affiliation{$^{1}$ Dipartimento di Chimica, Universit\`a  ``Sapienza'', P.le A. Moro 5, 00185 Roma, Italy} 
\affiliation{$^{2}$ Consorzio Interuniversitario per le Applicazioni di Supercalcolo Per Universit\`a e Ricerca (CASPUR), Via dei Tizii 6, 00185 Roma, Italy}
\affiliation{$^{3}$School of Physics, Room 302 UCD-EMSC, University College Dublin, Belfield, Dublin 4, Ireland}
\affiliation{$^{4}$Dipartimento di Fisica, Universit\`a di Cagliari, Cittadella Universitaria, I-09042 Monserrato (Ca), Italy}
\date{\today}

\begin{abstract}
We investigated the relative stability of the amorphous vs crystalline nanoparticles of size ranging between $0.8$ and $1.8$~nm. We found that, at variance from bulk systems, at low $T$ small nanoparticles are amorphous and they undergo to an amorphous-to-crystalline phase transition at high $T$. On the contrary, large nanoparticles recover the bulk-like behavior: crystalline at low $T$ and amorphous at high $T$. 
\textcolor{black}{We also investigated the structure of crystalline nanoparticles, providing evidence that they are formed by an ordered core surrounded by a disordered periphery. Furthermore, we also provide evidence that the details of the structure of the crystalline core depend on the size of the nanoparticle}

\end{abstract} 

\maketitle
\section{Introduction}
\label{sec:Introduction}
Nano-scale systems behave differently than ordinary bulk materials since, among other reasons, their physico-chemical properties do depend upon their size and shape. 
Considerable effort is ongoing to understand, design, fabricate, and manipulate materials at such a short length scale, so as to get tailored properties. In particular, the identification of how the structural features depend upon the actual thermodynamic conditions is attracting an increasing interest as it paves the way toward explaining the structure-property relationship, an issue of large technological impact. Among the nano-sized systems of current interest, silicon nano-particles embedded in amorphous SiO$_2$ are especially important for their possible application as photo-emitting materials for optoelectronics, \cite{daldosso_role_2003, science-292-258} as well as for the light harvesting component of solar cells.\cite{solarCells,solarCells2} 

A feature strongly affecting the properties of nano-sized semiconductor particles is whether they are crystalline or amorphous. In particular, it has been experimentally observed that the photoluminescence of Si nano-particles embedded in silica strongly depends (both in wavelength and intensity) on their crystallinity. Their structural evolution has been accordingly characterized: Si nano-particles are initially formed amorphous and then transformed into crystalline upon thermal annealing at high temperatures (typically at $1100 ^\circ$C or above). \cite{jap-95_3723, jap-103_114303, Wakayama1998124, jap-83_2228, Wang2006486, jp-19_225003} During annealing, another phenomenon has been nevertheless observed, namely: the growth of nano-particles, which makes it difficult to unambiguously identify the actual atomistic mechanisms driving the observed evolution. 
The aim of this paper is to resolve this ambiguity by elaborating a  thorough atomistic explanation of the observed microstructure evolution of an embedded Si nano-particle, through computer experiments addressed to measuring its free energy in different states of aggregation. The main output of the present investigation is that we identify the most stable phase of a Si quantum dot as a function of its size and the thermal conditions.   
This result provides evidence that at the nano-scale the relative stability of the ordered and disordered phases could be otherwise than in the bulk samples. In particular, we show that this result is able to explain the experimental findings on the mechanism of formation of crystalline nano-particles. \cite{jap-95_3723, jap-103_114303, Wakayama1998124, jap-83_2228, Wang2006486, jp-19_225003}
We also investigated the atomistic structure of the states corresponding to the minima of free energy, discovering that the ordered states are not simple crystal-like clusters; rather, they are made by a crystalline core surrounded by a disordered shell. \textcolor{black}{We also found that the details of the atomistic structure of the crystalline core depend on the size of the nanoparticle.} The analysis performed in this investigation is very general since it is addressed to any metastable state identified by atomistic simulations and to characterize the order-disorder phase change as a function of the size of the nanoparticle and the temperature of the system.

The article is organized as follows: in Sec. \ref{sec:TheoryAndSimulation} we describe the computational methods we used for the present free energy calculation and describe the simulation setup. In Sec. \ref{sec:ResultsAndDiscussion} we present and discuss our results. Finally, in Sec.  \ref{Sec:Conclusions} we draw some conclusions.

\section{Theoretical background, computational methods and simulation setup}
\label{sec:TheoryAndSimulation}
This section is divided into three Subsections. In Sec. (\ref{sec:Theory-methods}) we introduce the technique used to calculate the free energy of a computational sample in a given state. This technique requires the introduction of suitable collective variables describing the state of the system, which are described in full detail in  Sec. (\ref{sec:Theory-CollVarr}). Finally, in Sec. (\ref{sec:Theory-Setup}) we present the simulation setup. 

\subsection{Free energy calculation}
\label{sec:Theory-methods}
In this work we compute the free energy by numerically integrating the gradient of the free energy, which is evaluated according to the restraint method introduced by Maragliano and Vanden-Eijnden. \cite{Maragliano2006168} This approach allows to efficiently compute the (relative) free energy in a system containing multiple metastable states separated by significant free energy barriers. 

Let $\theta(\mathbf{x})$ be a suitable order parameter describing the state of the system, where $\mathbf{x}$ is the $3 N$ vector of the atomic positions. Consider the following Hamiltonian:

\begin{eqnarray}
H_k(\mathbf{p}, \mathbf{x}) = K(\mathbf{p}) + V(\mathbf{x}) + {k \over 2} (\theta(\mathbf{x}) - \theta^*)^2
\end{eqnarray}

\noindent where $\mathbf{p}$ is the $3 N$ vector of the atomic momenta, $K(\mathbf{p})$ is the kinetic energy, and $V(\mathbf{x})$ is the interatomic potential. $\theta^*$ is a possible realization of the collective variable $\theta(\mathbf{x})$ and $k$ is a tunable parameter. Below, we show that in the limit $k \rightarrow \infty$ the observable $<k (\theta(\mathbf{x}) - \theta^*)>_{H_k}$, where $<\cdots>_{H_k}$ indicates the average over the canonical ensemble associated to the Hamiltonian $H_k(\mathbf{p}, \mathbf{x})$, 
is the derivative of the free energy with respect to the parameter $\theta$ identifying the state of the system at the value $\theta=\theta^*$ ($d F(\theta)/ d \theta |_{\theta = \theta^*}$). By simple algebra it can be shown that 

\begin{eqnarray}
\label{eq:gradFEstimator}
<k (\theta(\mathbf{x}) - \theta^*)>_{H_k} && = \\ 
- {d \over d \theta^*} &&{\beta^{-1} \log \int d\mathbf{x} d\mathbf{p} \exp[-\beta H_k(\mathbf{x}, \mathbf{p})] \over {\mathcal Z} } \nonumber
\end{eqnarray}

\noindent where ${\mathcal Z} = \int d\mathbf{x} d\mathbf{p} \exp[-\beta H(\mathbf{x}, \mathbf{p})]$ is the canonical partition function of the system and $\beta = k_B T$ ($k_B$ is the Boltzmann constant). 
Since $\mathcal Z$ is $\theta$-independent
its introduction does not affect our argument but it is necessary for the probabilistic interpretation of $<k (\theta(\mathbf{x}) - \theta^*)>_{H_k}$.
Let us now consider  $ \int d\mathbf{x} d\mathbf{p} \exp[-\beta H_k(\mathbf{x}, \mathbf{p})] / {\mathcal Z}$ in the limit of large $k$:

\begin{eqnarray}
\label{eq:PDFLargeKLimit}
\lim_{k \rightarrow \infty} && {\int d\mathbf{x} d\mathbf{p} \exp[-\beta H_k(\mathbf{x}, \mathbf{p})] \over {\mathcal Z} }  =  \\
&&{ \int d\mathbf{x} d\mathbf{p} \exp[-\beta H(\mathbf{x}, \mathbf{p})] \delta(\theta(\mathbf{x}) - \theta^*) \over {\mathcal Z} }  \nonumber
\end{eqnarray}

\noindent The r.h.s of Eq.~\ref{eq:PDFLargeKLimit} is, by definition, the probability density $P_\theta(\theta^*)$ to find the system in a state corresponding to $\theta(\mathbf{x}) = \theta^*$. The relation between the free energy $F(\theta^*)$ and the above probability density function is $F(\theta^*) = -\beta^{-1} \log P_\theta(\theta^*)$. As a result, in the limit mentioned above,  Eq.~\ref{eq:gradFEstimator} reads:

\begin{equation}
\lim_{k \rightarrow \infty} <k (\theta(\mathbf{x}) - \theta^*)>_{H_k} = d  F(\theta) / d \theta|_{\theta = \theta^*} 
\end{equation}

\noindent By (numerically) integrating the so computed $d  F(\theta) / d \theta|_{\theta = \theta^*}$ we straightforwardly get the $F(\theta)$ curve.  

In practice, we can compute an approximation to the derivative of the free energy on the collective variable $\theta$ by calculating $<k (\theta(\mathbf{x}) - \theta^*)>_{H_k}$ for large $k$ by Molecular Dynamics (MD). In this case, we replace the ensemble average $<k (\theta(\mathbf{x}) - \theta^*)>_{H_k}$ by a time average of the operator $k (\theta(\mathbf{x}(t)) - \theta^*)$ along the trajectory of a constant temperature MD in which the atomic forces are obtained from the potential $U(\mathbf{x}) = V(\mathbf{x}) + k/2 (\theta(\mathbf{x}) - \theta^*)^2$. 
Other methods could also be used to compute the free energy (e.g. the Blue Moon ensemble \cite{Carter1989} or the umbrella sampling \cite{Valleau1977}). The advantage of the method used in this paper with respect to the Blue Moon sampling is that it does not require any un-biasing, which might be difficult to perform, depending on the selected collective variable; on the other hand, with respect to the umbrella sampling the advantage is that it does not require a technique such as the Weighted Histogram Analysis Method (WHAM) \cite{WHAM} to reconstruct the free energy curve; rather, this quantity is obtained by performing a simpler numerical integration. Nevertheless, the accurate calculation of the free energy $F(\theta)$ might require the calculation of  $<k (\theta(\mathbf{x}) - \theta^*)>_{H_k}$ in more $\theta^*$ points in comparison to the corresponding number of positions of the umbrella potential (i.e. number of umbrella sampling runs). It is worth to remark that the present method allows to compute a free energy including both the configurational and vibrational entropic contributions with no approximation (apart those connected with finite time simulation).

In the present investigation we need to extend the approach described above to the case of two collective variables, one  controlling the degree of order of the nanoparticle (and, therefore, the amorphous vs crystalline phase), the other monitoring its size. They are both described in detail in Sec. \ref{sec:Theory-CollVarr}. The use of two collective variables is motivated by the need of computing the relative free energy of the ordered and disordered phase at a given size of the nanoparticle. The probability density function associated to such a free energy is the conditional probability density function to observe $\theta = \theta^*$ (the first collective variable) given $\phi = \phi^*$ (the second one), hereafter indicated as $P(\theta^*| \phi^*)$. This quantity can be computed by numerically integrating the observable $<k (\theta(\mathbf{x}) - \theta^*)>_{H_{k,k'}}$, where the Hamiltonian $H_{k,k'} = K(\mathbf{p}) + V(\mathbf{x}) + {k \over 2} (\theta(\mathbf{x}) - \theta^*)^2 + {k' \over 2} (\phi(\mathbf{x}) - \phi^*)^2$. As explained before, this observable can be computed by MD.

The procedure described above assumes that, apart from the collective variable $\theta(\mathbf{x})$ and $\phi(\mathbf{x})$, the system is ergodic. However, there might be cases in which other slow variables are present in the system and, therefore, the calculation of the observable $<k (\theta(\mathbf{x}) - \theta^*)>_{H_{k,k'}}$ cannot be efficiently computed by a straightforward MD simulation. In Sec. \ref{sec:ResultsAndDiscussion} we show that this is indeed the case in the present investigation. In order to overcome this problem we combine the restrained MD method described above with the Replica Exchange method. \cite{parallelTempering} A similar approach has already been used in the simulation of rare events in which the replica exchange method has been used in combination with the umbrella sampling. \cite{Coluzza1953} or Metadynamics \cite{Bussi2006} The Replica Exchange technique consists in running several MD simulations at different temperatures in parallel and, from time to time, to swap  the current microstate (i.e. the instantaneous set of atomic positions and momenta) between two parallel runs. The swapping is accepted/rejected according to a Metropolis Monte Carlo criterion, namely with the probability

\begin{eqnarray}
p = && \min \big \{1, \\
&& \exp[(V_{k,k'}(\{\mathbf{x}, \mathbf{p}\}_{\beta_i})-V_{k,k'}(\{\mathbf{x}, \mathbf{p}\}_{\beta_j}))(\beta_i - \beta_j) \big \} \nonumber 
\end{eqnarray}

\noindent where $V_{k,k'}(\{\mathbf{x}, \mathbf{p}\}_{\beta_i})$ and $V_{k,k'}(\{\mathbf{x}, \mathbf{p}\}_{\beta_j})$ are the potential energies of the two microstates, respectively at $\beta_i = k_B T_i$ and $\beta_j = k_B T_j$ in the phase space points $\{\mathbf{x}, \mathbf{p}\}_{\beta_i}$ and $\{\mathbf{x}, \mathbf{p}\}_{\beta_j}$ at the moment of the attempted swapping. If the swap is accepted, then the microstates corresponding to temperatures $T_i$ and $T_j$ are simply interchanged. If the swap is rejected, the microstates are further aged at their own temperature. 
The key feature of this method is that the sampling of the system phase space obtained by the piece-like replica exchange trajectories is consistent with the canonical (conditional) probability density function at each target temperature. However, since the individual pieces of the replica exchange trajectories are obtained by swapping from higher temperatures, they more likely overcome possible free energy barriers. In short: the replica exchange trajectories are ergodic. 

\textcolor{black}{The simulation techniques described above (and the collective variables presented in the next section) have been implemented in the CMPTool simulation package. \cite{CMPTool, reordering, reorderingAndQuickCells} In particular, the combination of the restrained MD with Parallel Replica technique allows a two-level parallel approach. The fist level of parallelism is over the replicas while the second, implemented according to the usual domain decomposition paradigm, is within each replica. While the latter level of parallelism, that requires a tight connection, was implemented using the Message Passing Interface (MPI) application programming interface, the former was implemented at a scripting level. This allowed to run the simulations on a cluster of loosely connected multicore machines communicating by a Gigabit network.}

\subsection{Collective variables for modeling the crystallization process in nanoparticles} 
\label{sec:Theory-CollVarr}
We now discuss the collective variables used to study the crystallization process. It is worth to stress again that in this paper we are not interested in studying the detailed mechanism of crystallization in nanoparticles. Rather, we investigate the ``relative stability'' of the ordered vs the disordered phase as a function of the size of the nanoparicles and the temperature. Therefore, our collective variables must be able to distinguish between the crystalline and the amorphous phase (i.e. they need to be order parameters), rather than modeling the mechanism of the crystallization. 


\subsubsection{${\mathcal R}({\mathbf x})$ order parameter to monitor the size of the nanoparticle}
\label{sec:Theory-CollVarr-Size}
We introduce the notion of size of the nanoparticle by a collective variable denoted by the symbol ${\mathcal R}({\mathbf x})$. In a system in which the nanoparticle is made of atoms of type `A' (Si in this case) and the matrix is made, or contains, atoms of type `B' (O in this case) a possible definition of the collective coordinate ${\mathcal R}({\mathbf x})$ is the distance between the center $x_c$ of the nano-particle (a point kept fixed during the simulations) and the closest oxygen atom, i.e. ${\mathcal R}({\mathbf x})=\min_i \{|x_c -  x^O_i|\}$, where $x^O_i$ is the coordinate of the $i$-th oxygen atom. The force acting on the atoms associated to this collective variable cannot be straightforwardly evaluated since ${\mathcal R}({\mathbf x})$ is a non-analytical function of ${\mathbf x}$ and, therefore, there is no way to proceed through the direct calculation of $\nabla k'/2 ({\mathcal R}({\mathbf x}) - {\mathcal R}^*)^2$. We replaced the above definition of the collective variable by a smooth analytical approximation of it that, in a proper limit, converges to its exact definition and performed biased MD runs according to this representation of ${\mathcal R}({\mathbf x})$. This smooth analytical approximation is obtained in two steps: (i) first we obtain an analytic (explicit) expression of $\min_i \{|x_c -  x^O_i|\}$ as a function of the positions $x^O_i$, and (ii) then we introduce a smooth approximation to this expression. The first step consists in recognizing the following identity:

\begin{eqnarray}
\label{eq:Ranalytical}
\min_i\{|{\mathbf x}_c -  {\mathbf x}^O_i|\} &\equiv& \sum_{i}  \Big [ |{\mathbf x}_c -  {\mathbf x}^O_i|  \nonumber \\
 && \prod_{j \neq i}^{N_i} \Theta(|{\mathbf x}_c -  {\mathbf x}^O_j| - |{\mathbf x}_c -  {\mathbf x}^O_i|)  \Big ]
\end{eqnarray}

\noindent where $\Theta({\mathbf x})$ is the Heaviside step function. Let $l$ be the $O$ atom closest to the center of the nano-particle, then $  \prod_{j \neq i}^{N_i} \Theta(|{\mathbf x}_c -  {\mathbf x}^O_i| - |{\mathbf x}_c -  {\mathbf x}^O_j|) = \delta_{il}$  , where $\delta_{il}$ is the Kronecker symbol. 
This implies that the result of this sum is $ |{\mathbf x}_c -  {\mathbf x}^O_l|$, i.e. the distance from ${\mathbf x}_c$ of the closest O atom. Eq.~\ref{eq:Ranalytical} is, therefore, the analytical expression of the collective coordinate ${\mathcal R}({\mathbf x})$.  A smooth approximation to ${\mathcal R}({\mathbf x})$ can be obtained by replacing the Heaviside step function by a sigmoid function. We used a sigmoid function expressed in term of the Fermi function:   
 
\begin{equation}
S(t) = 1 - {1 \over \ 1 + \exp[\lambda t]}
\end{equation}

\noindent where $\lambda$ is the parameter controlling its smoothness. In our simulations $\lambda$ has been chosen such that the sigmoid function goes from $0.95$ to $0.05$ in one atomic layer ($\sim 0.2$~nm). A consequence of this replacement is that the size of the nano-particle is now defined as a weighted average of the distance of one atomic layer of oxygen atoms from the centre of the nano-particle.

\subsubsection{${\mathcal Q}_6({\mathbf x})$ order parameter to monitor the phase of the nanoparticle}

We compute the free energy of a Si nano-particle embedded in silica as a function of its degree of order, as measured by the bond-orientational order parameter (${\mathcal Q}_6({\mathbf x})$) introduced by Steinhardt et al. \cite{PhysRevB.28.784} for bulk systems. In this paper we adapted the original definition to the case of confined systems, as described below in detail. 

In general, ${\mathcal Q}_6({\mathbf x})$ is defined as

\begin{equation}
\label{eq:AppQlTot}
{\mathcal Q}_6({\bf x}) = \left ( {4 \pi \over 2 \times 6 + 1}  \sum_{m=-6}^6 |{\mathcal Q}_{6m}({\bf x})|^2 \right )^{1 \over 2}
\end{equation}

\noindent where ${\mathcal Q}_{6m}({\bf x})$ is the normalized and weighted sum of atomic vectors ${\it q}^i_{6m}({\bf x})$ (defined below) which, in bulk systems, reads

\begin{equation}
\label{eq:AppQlm}
{\mathcal Q}_{6m}({\bf x}) =  {\sum_{i=1}^{N} N_i {\it q}^i_{6m}({\bf x}) \over \sum_{i=1}^N N_i}
\end{equation}

\noindent where $N$ is the number of atoms in the system, $N_i$ is the number of nearest neighbors of the atom $i$ and $m= -6, \dots, 6$. In the case of confined systems we limit the sum over $i$ to just the atoms belonging to the nano-particle. Consistently with our definition of the size of the nanoparticle, we identify these atoms as those at a distance lower than ${\mathcal R}^*$ from the center of the nanoparticle (${\mathcal R}^*$ is the size of the nanoparticle, see Sec. \ref{sec:Theory-CollVarr-Size}). According to this definition, the ${\mathcal Q}_6m({\mathbf x})$ of the nanoparticle is: 

\begin{equation}
\label{eq:AppQlmNano}
{\mathcal Q}_6m({\bf x}) = \frac{ \sum_{i=1}^N N_i {\it q}_{6m}^i({\bf x}) \left ( 1 - \Theta(|{\bf x}_i^{Si} - {\bf x}_c| - {\mathcal R}^*) \right )}{\sum_{i=1}^N N_i}
\end{equation}

\noindent where $\Theta(x)$ is the Heaviside step function. As for the case of the collective variable ${\mathcal R}({\mathbf x})$, the Heaviside step function is, in practice, replaced by a sigmoid function  $S(t) = 1 - {1 / (1 + \exp[\lambda t])}$. 

The ${\it q}^i_{6m}(x)$ function appearing in Eq.~\ref{eq:AppQlmNano} is defined according to the following expression:

\begin{equation}
{\it q}^i_{6m}({\bf x}) =  {\sum_{j=1}^{N_i} {\it Y}_{6m}({ \hat x_{ij}}) \over N_i}
\end{equation}

\noindent where $ {\it Y}_{6m}({ \hat x_{ij}})$ is the spherical harmonic function of degree $6$ and the component $m$ computed on the solid angle $\hat x_{ij}$ formed by the distance vector ${\vec x}_{ij}$ and the reference system. The sum runs over the $N_i$ nearest neighbors of the atom $i$. 

The sum over the $m$ component in Eq.~\ref{eq:AppQlTot} makes the collective coordinate ${\mathcal Q}_6({\bf x})$ rotationally invariant, i.e. independent on the orientation of the reference system. 

When the system is an ideal crystal and the temperature is $0$~K, the environment of all the atoms is the same and, therefore, ${\mathcal Q}_{6}({\bf x})$ is maximum as there is not interference among the ${\it q}^i_{6m}({\bf x})$. On the contrary, in a perfectly disordered system the orientation of bonds is random and, therefore, there is complete interference among the ${\it q}^i_{6m}({\bf x})$, and ${\mathcal Q}_{6}({\bf x})$ is zero. However, even when the system is at finite temperature and its size is finite, this order parameter is still able to distinguish between a disordered and a crystalline phase. 

Before concluding this section it is worth to mention that the ${\mathcal Q}_{6}$ collective variable, or its modifications, has been already used to study crystallization by atomistic simulations \cite{ReintenWolde1996, Auer2001,PhysRevLett.94.235703} and experiments. \cite{Gasser2001}

%

\subsection{Simulation setup} 
\label{sec:Theory-Setup}
In the present investigation, the restrained MD is governed by the superposition of the physical potential  and the restraining potential $k/2({\mathcal Q}_6({\mathbf x}) - {\mathcal Q}_6^*)^2$ $+ k'/2({\mathcal R}({\mathbf x}) - {\mathcal R}^*)^2$. $k$ and $k'$ are the parameters controlling the degree of biasing and must be large enough such that along the MD the values of ${\mathcal Q}_6({\mathbf x})$ and ${\mathcal R}({\mathbf x})$ oscillate about the target values ${\mathcal Q}_6^*$ and ${\mathcal R}^*$, respectively. \textcolor{red}{In this work we use the Billeter et al. \cite{PhysRevB.73.155329} environment-dependent force field. This classical force field, which is an extension of the Tersoff potential \cite{tersoff}, is defined as the sum of generalized Morse pair potential terms: $V({\bf x}) = \sum_{i>j} v_{IJ}(|{\bf x}_i - {\bf x}_j|)$, where $I$ and $J$ denote the chemical species of atoms $i$ and $j$, respectively. The pair potential $v_{IJ}(x)$ is given by  $v_{IJ}(x) = f_{IJ}(x) [A_{IJ} \exp[-\lambda_{IJ} x] - b_{IJ}({\bf x})A_{IJ} \exp[-\mu_{IJ} x]] $. $f_{IJ}(x)$ is a switching function that makes $v_{IJ}(x)$ to go smoothly to zero at the cutoff distance $r_{ij}^{cut}$ ($r_{ij}^{cut}$ is such that $v_{IJ}(x)$ is non zero only between nearest neighbor atoms). $b_{IJ}({\bf x})$ is the environment dependent term, which is function of an effective coordination number. The effective coordination number embodies three-body terms through the angle formed by $i$, $j$ and all their nearest neighbors. In addition to the terms mentioned above, the Billeter et al. force field contain terms that 
prevent unphysical over/under-coordination at the Si/SiO$_2$ interface. For a detailed description of the potential we refer the reader to the original paper.}  The reliability of this force field in modeling equilibrium and dynamical properties of Si nano-particles embedded in silica, of the Si/a-SiO$_2$ interface and Si nanowires has been already established. \cite{PhysRevB.73.155329, fischer:012101,ippolito:153109, PhysRevB.81.014203, tuma:193106}

The computational samples are prepared by thermally annealing a periodically-repeated amorphous silica system and embedding Si nano-grains (extracted from a well equilibrates either amorphous or crystalline bulk). The amorphous silica sample is prepared through the quenching-from-the-melt procedure, that is by cooling down very slowly a high temperature SiO$_2$ melt. The total system contains from $\sim 6000$ to $\sim 12000$ particles, corresponding to a nano-particle radius varying in the range $1-2$~nm. Computational samples are first thermalized at $300$~K \textcolor{red}{and ambient pressure by using the Martyna-Tobias-Klein variable cell algorithm\cite{MTK}} in order to release possible stress at the Si/silica interface. Typically, during such a thermalization step, the nano-particles slightly shrink. After this initial step, we impose the restraint on the size of the nano-particles and thermalize the samples at the various target temperatures. Because of the restraint on their size, at this stage  the size of the nano-particles does not change. After this treatment the samples are ready for the restrained simulations described above. \textcolor{red}{The simulations are performed at a fixed volume. However, we checked that the pressure is close to the ambient value all along the simulation.}

In order to verify possible artifacts due to finite-size effects, we repeated the calculation of the observable $<k (\theta(\mathbf{x}) - \theta^*)>_{H_{k,k'}}$ at few selected value of ${\mathcal Q}_6^*$ and ${\mathcal R}^*$ on samples of different size of the silica matrix. We did not observe any significant difference in them (the differences were within the statistical error). This demonstrates that there are no finite-size effects in our free energy calculations. 

\textcolor{red}{We compute the free energy $F({\mathcal Q}_6, {\mathcal R})$ in the range ${\mathcal Q}_6 \in [0, 0.65]$ and ${\mathcal R} \in [0.8 , 1.8]$~nm. 
The rationale for the upper limit of the ${\mathcal Q}_6$ range is that the value of the bond-orientational order parameter for an ideal Si crystal is ${\mathcal Q}_6 = 0.63$ and, since from experiments and previous calculations it is known that Si crystalline nanoparticles assume a structure with a (distorted) diamond-like core and a disordered periphery \cite{daldosso_role_2003,Wakayama1998124,jap-83_2228,PhysRevLett.93.226104}, we expect the ${\mathcal Q}_6$ of crystalline nanoparticles be lower than this limit. The samples created according to the protocol described above confirm that the ${\mathcal Q}_6$ of crystalline nanoparticles is lower  than this upper limit. However, after a preliminary scanning of $\nabla_{{\mathcal Q}_6 } F({\mathcal Q}_6, {\mathcal R})$ that allowed to identify the region of ${\mathcal Q}_6 $ containing the minima in the above domain, we restricted the calculations to a smaller range: $[0.04, 0.285]$, $[0.07, 0.35]$ and $[0.03, 0.38]$ for the nanoparticles of radius $0.8$, $1.3$ and $1.8$~nm, respectively.
As for the size of nanoparticles, the same experiments mentioned above report unusual phase transitions (disorder-to-order with growing $T$) for nanoparticles in the range $[0.5, 2.0]$~nm. We decided therefore to study nanoparticles of three size in this range: $0.8$, $1.3$ and $1.8$~nm.} 

\textcolor{red}{Finally, the restrained MD/parallel tempering simulations where performed at $300$, $500$, $750$, $1000$, $1250$ $1500$ and $1750$~K. }

\section{Results and discussion}
\label{sec:ResultsAndDiscussion}
\begin{figure}
 \begin{center}
 \includegraphics[width=0.45\textwidth]{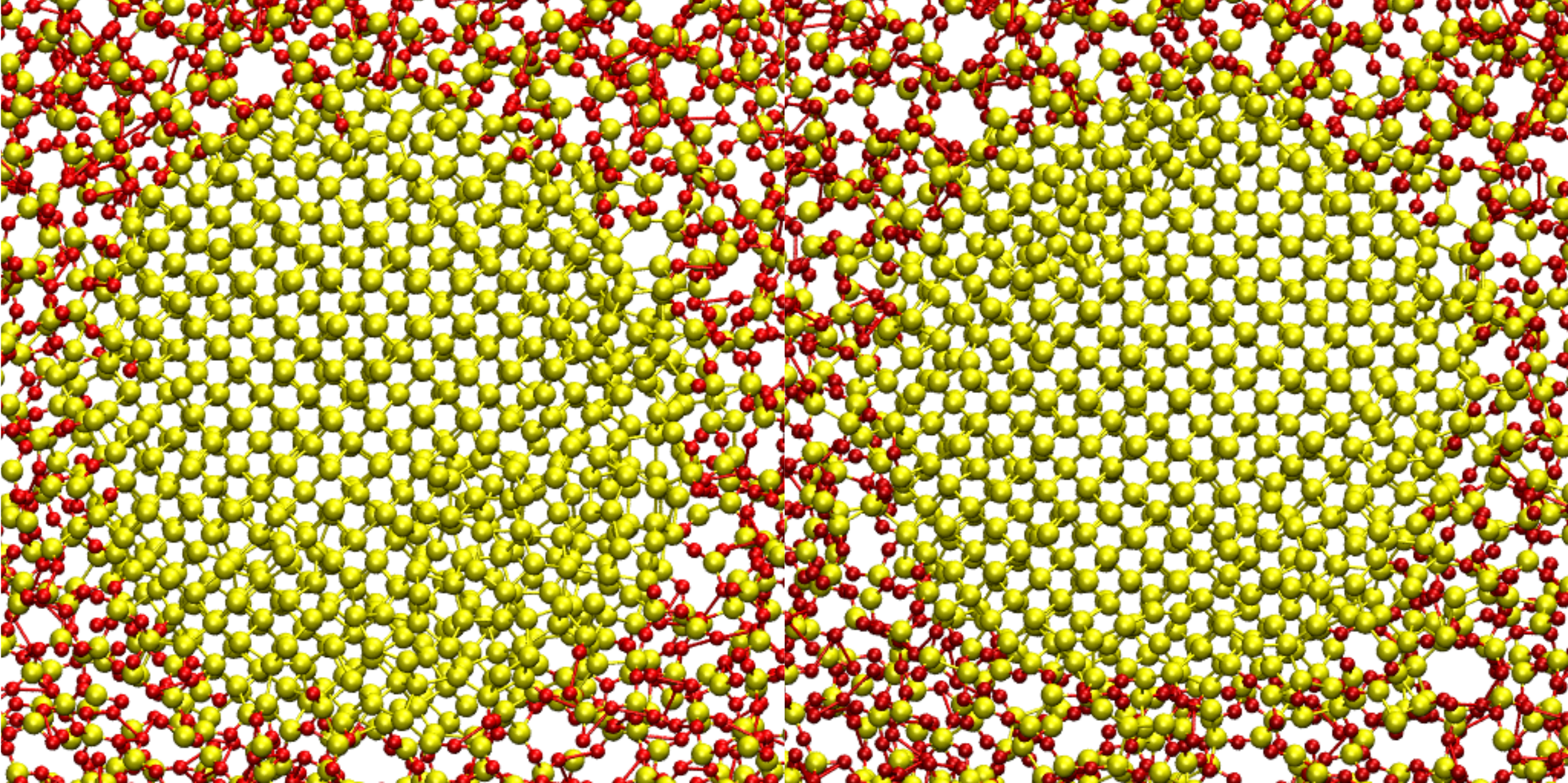}
 \end{center}
 \caption{Two different configurations of an embedded silicon nano-particle with radius as large as $1.8$~nm. They both correspond to ${\mathcal Q}_6^* = 0.19$. Oxygen atoms are displayed in red and Silicon atoms in yellow. In order to improve the readability, only the atoms laying within a $1.5$~nm-thick slice are drawn.}\label{defectSnapshots}
 \end{figure}

As a first preliminary step, 
we begin the presentation and discussion of our results by motivating
the use of the restraint method in combination with the replica exchange method through a relevant example. In Fig.~\ref{defectSnapshots} two quasi-crystalline configurations are shown, corresponding to the same value of the ${\mathcal Q}_6$ parameter. These configurations embed different defected structures. The configuration shown in the left panel is characterized by an extended disordered region in the bottom-right part of the nano-particle. At a variance, two smaller disordered regions characterize the second configuration shown in right panel, respectively in the bottom-right and top-left part of the nano-particle. Both configurations should be considered for the correct evaluation of the gradient of the free energy $<k (\theta(\mathbf{x}) - \theta^*)>_{H_{k,k'}}$ corresponding to the same value of ${\mathcal Q}_6^*$. However, a sizable free energy barrier likely separates them. Therefore, if the simulation is started from one of the two configurations, then the second one likely could not be visited during the explored time scale. By using the replica exchange method we are able to properly and efficiently sample both of them and compute $d  F(\theta) / d \theta$ at the present ${\mathcal Q}_6^*$ accurately.

As a second preliminary step, we summarize the experimental findings we aim to address. By comparing Energy Filtered Transmission Electron Microscopy (EFTEM) and Dark-Field Transmission Electron Microscopy (DFTEM) images in Si-rich SiO$_x$ samples it was shown that Si nano-particles start to form at $1000 ^\circ$C. \cite{jp-19_225003} At this temperature they are all amorphous, while at $1100 ^\circ$C about one third become crystalline. By further increasing the annealing temperature by $50 ^\circ$C, the fraction of crystalline nano-particles rises up to ~60\%, while the average size of the nano-particles and the distribution of their size remains almost unchanged. Finally, at the annealing temperature of $1250 ^\circ$C 100\% of nano-particles are crystalline. At this temperature the average size is slightly increased, but the particle size distribution is still largely superimposed to the distributions observed at $1100 ^\circ$C and $1150 ^\circ$C. It was also found that the system is stationary with respect to the amorphous~vs.~crystalline composition. Similar investigations have been performed on Si/SiO$_2$ multilayers \cite{jp-19_225003} where the growth of the crystalline fraction with the annealing temperature is even more sudden: the degree of crystallinity increases from about 15\% to 90\% when the annealing temperature is increased from $1100 ^\circ$C to $1200 ^\circ$C. Also in this case it was demonstrated that the samples are at the equilibrium.

\begin{figure*}
  \begin{center}
\includegraphics[width=0.85\textwidth]{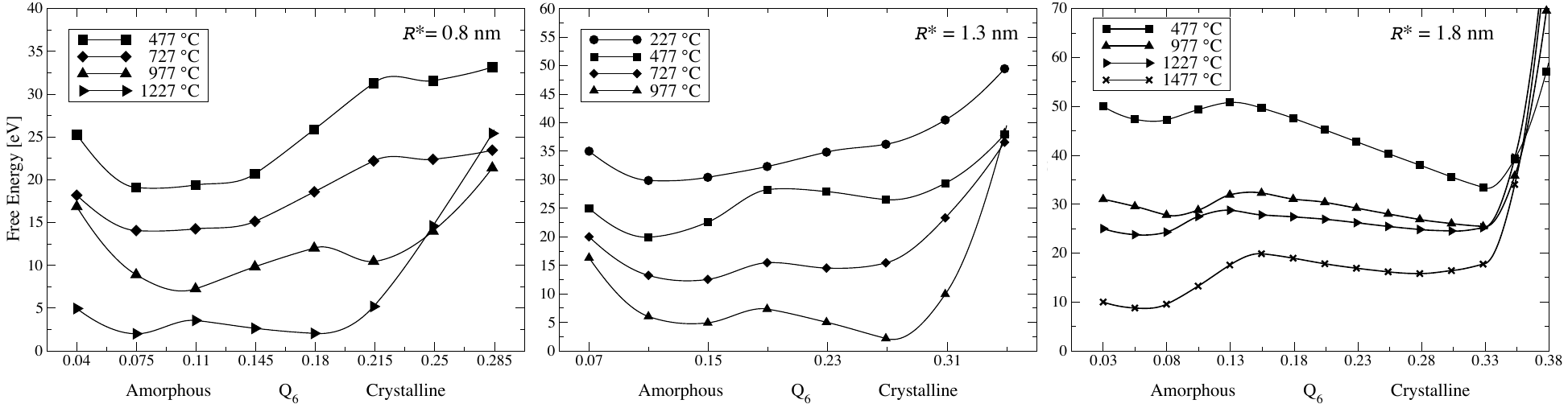}
 \end{center}
 \caption{Free energy vs ${\mathcal Q}_6$ curves for nano-particles with radius $0.8$~nm, $1.3$~nm and $1.8$~nm. The curves are shifted to improve readability.}\label{FOfQ6}
 \end{figure*}

\begin{figure*}
 \begin{center}
 \includegraphics[width=0.90\textwidth]{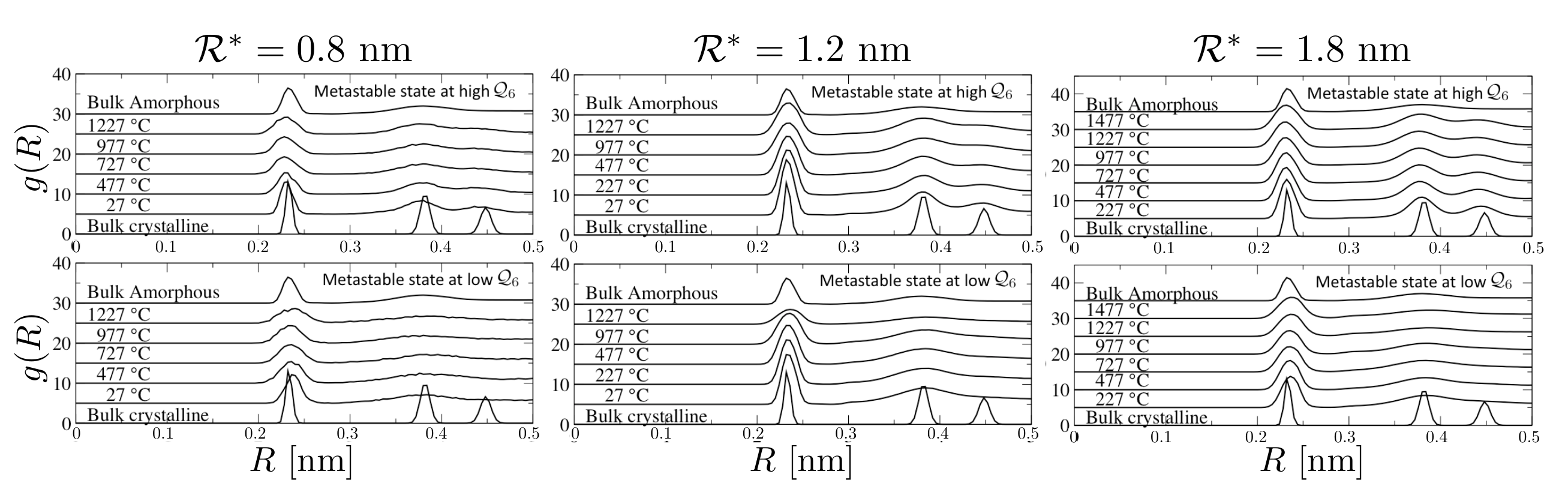}
 \end{center}
 \caption{Si-Si pair correlation function ($g(R)$) of the low and high ${\mathcal Q}_6$ metastable states at various $T$. Bulk crystalline and amorphous $g(R)$ are also reported for comparison.}\label{fig:GRVsT}
 \end{figure*}

We can now turn to the results of our simulations. In Fig.~\ref{FOfQ6} the free energy curves $F({\mathcal Q}_6^*)$ of Si nano-particles of size ${\mathcal R}^* = 0.8$~nm, ${\mathcal R}^*= 1.3$~nm, and ${\mathcal R}^* = 1.8$~nm at various temperatures in the range $227 ^\circ$C - $1477 ^\circ$C are shown, as obtained from our simulations (our calculations are performed in Kelvin, while the results are presented in Celsius for homogeneity with available experimental data). \textcolor{red}{Above and throughout the paper we shall denote the radius of the nanoparticle by the symbol ${\mathcal R}^*$, which is the target value of the collective variable ${\mathcal R}({\bf x})$ and, via the corresponding restrain term (see Sec. \ref{sec:Theory-Setup}), defines its size.}
As a first remark, we notice that two metastable states, one a low ${\mathcal Q}_6$ and one at higher  ${\mathcal Q}_6$, are present at all temperatures and sizes. In general, to high values of  ${\mathcal Q}_6$ correspond crystalline states while to low values of the same order parameter correspond disordered (amorphous) states. However, especially for the smallest nanoparticle, the difference between the value of  ${\mathcal Q}_6$ corresponding to the two metastable states is small. Therefore, the identification of the phase corresponding to a given state requires a further investigation of the structure. We performed this analysis by computing the Si-Si partial pair correlation function $g(R)$ for the atoms belonging to the nanoparticle ($|{\mathbf x}^{Si}_i - {\mathbf x}_c| \leq {\mathcal R}^*$) on the configurations corresponding to the two metastable states (${\mathcal Q}_6({\mathbf x}) = {\mathcal Q}_6^*$, where ${\mathcal Q}_6^*$ corresponds to one of the two minima). In the left-most panel of Fig.~\ref{fig:GRVsT} we report the $g(R)$ of the low (bottom panel) and high (top panel) metastable states for the ${\mathcal R} = 0.8$~nm nanoparticle at various temperatures. For the sake of comparison, we also show the $g(R)$ of the bulk amorphous and crystalline states at $T = 627 ^\circ C$ (i.e. $T = 900 K$). 
For the high ${\mathcal Q}_6$ metastable state, 
we notice that even at the highest $T$ the $g(R)$ is characterized by three peaks in the range $0 \leq r \leq 5$. These peaks correspond to the bulk-like first, second, and third neighboring shell, respectivelly. 
They broaden and reduce in height by increasing the temperature; nevertheless, they are still well visible even at $T = 1227 ^\circ C$. In general, even at low $T$, these peaks are broader than the corresponding bulk crystalline ones. This is due to a structure of the nanoparticle which is composed of two regions : a crystal-like core and a disordered surrounding shell. The latter is in contact with the SiO$_2$ matrix. This two-region structure is consistent with the structure found by Hadjisavvas and Kelires \cite{PhysRevLett.93.226104} in their investigation on crystalline nanoparticles embedded in silica. We shall provide further evidence of this two-region structure below. 

Let us move to the analysis of the structure of the low ${\mathcal Q}_6$ metastable state. As a first remark we notice than in the same $R$ range analyzed for the high ${\mathcal Q}_6$ case there are only two peaks. The first one, sharp and intense, corresponds to the nearest neighbor Si-Si pairs. The second one, very broad and small, is usually assigned to the second neighbor pairs, which in amorphous system are distributed over a large $r$ range. There is no other peak in the  $0 \leq r \leq 5$ domain. Comparing the $g(R)$ of the low ${\mathcal Q}_6$ metastable state of this particle with amorphous bulk Si we notice that there is a one to one correspondence between the equivalent peaks in the two system. Similar results, both for the low and high ${\mathcal Q}_6$ metastable states, are found for the ${\mathcal R} = 1.3$~nm and ${\mathcal R} = 1.8$~nm nanoparticles (see the central and right-most panels of Figs.~\ref{fig:GRVsT}, respectively). On the basis of this analysis, we identified the high ${\mathcal Q}_6$ metastable state of all the nanoparticles at all temperatures to be of crystalline nature while the one at low ${\mathcal Q}_6$ to be of amorphous type.

\begin{figure*}
 \includegraphics[width=0.98\textwidth]{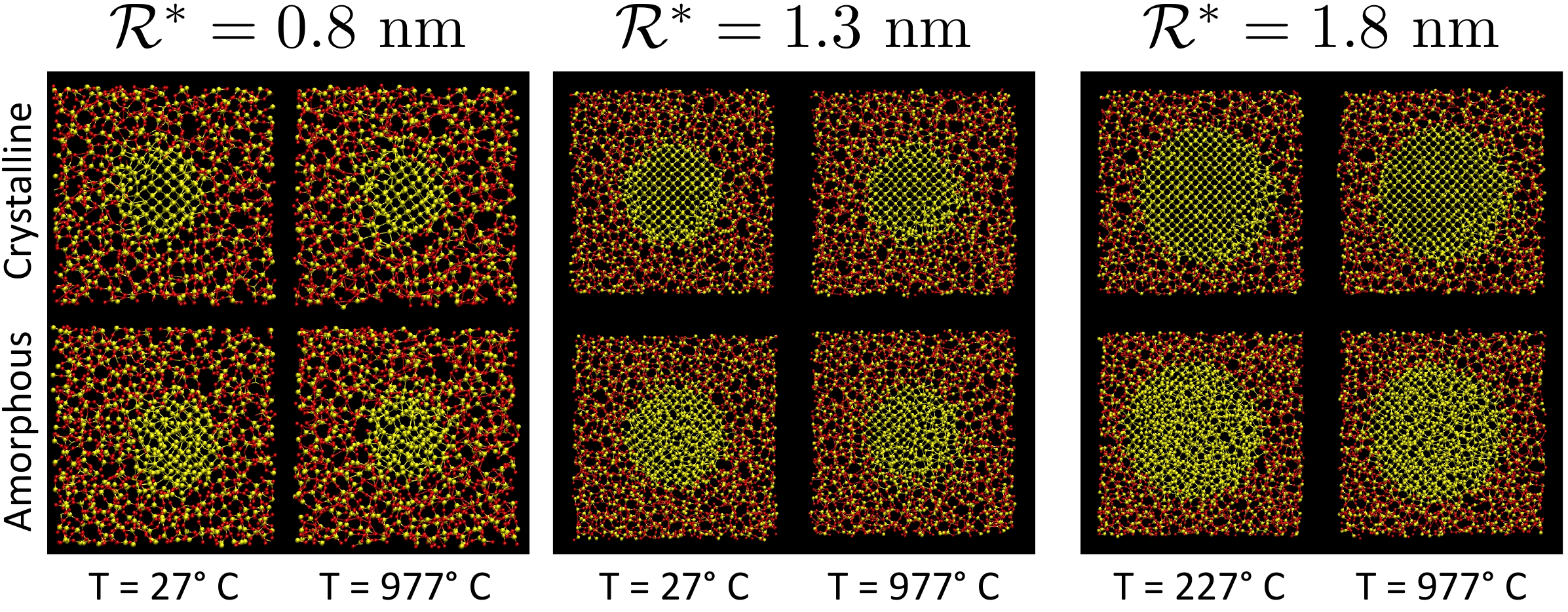}
 \caption{(color online) Snapshots of the nanoparticle of various size in the amorphous and crystalline metastable states at low and high temperature. Light (yellow) and dark (red) spheres represent the Si and O atoms, respectivelly.}\label{fig:snapshots}
 \end{figure*}

\textcolor{black}{The above conclusions are confirmed by the visual inspection of the structure of the nanoparticles in the low and high ${\mathcal Q}_6$ domains. In Fig.~\ref{fig:snapshots} we show few snapshots collected along the restrained MD at the values of the ${\mathcal Q}_6$ collective variable corresponding to the minimum of the free energy in the amorphous and crystalline metastable states, respectivelly, at low and high temperatures. It appears evident that the nanoparticles in the high ${\mathcal Q}_6$ domain have a crystal-like core in which the tetrahedral orientation of the Si-Si bonds is preserved. This core is surrounded by a shell containing disordered regions. 
At a variance from this, the structure of the nanoparticles corresponding to the low ${\mathcal Q}_6$ free energy minimum are completely disordered.}

\begin{figure}
 \begin{center}
 \includegraphics[width=0.45\textwidth]{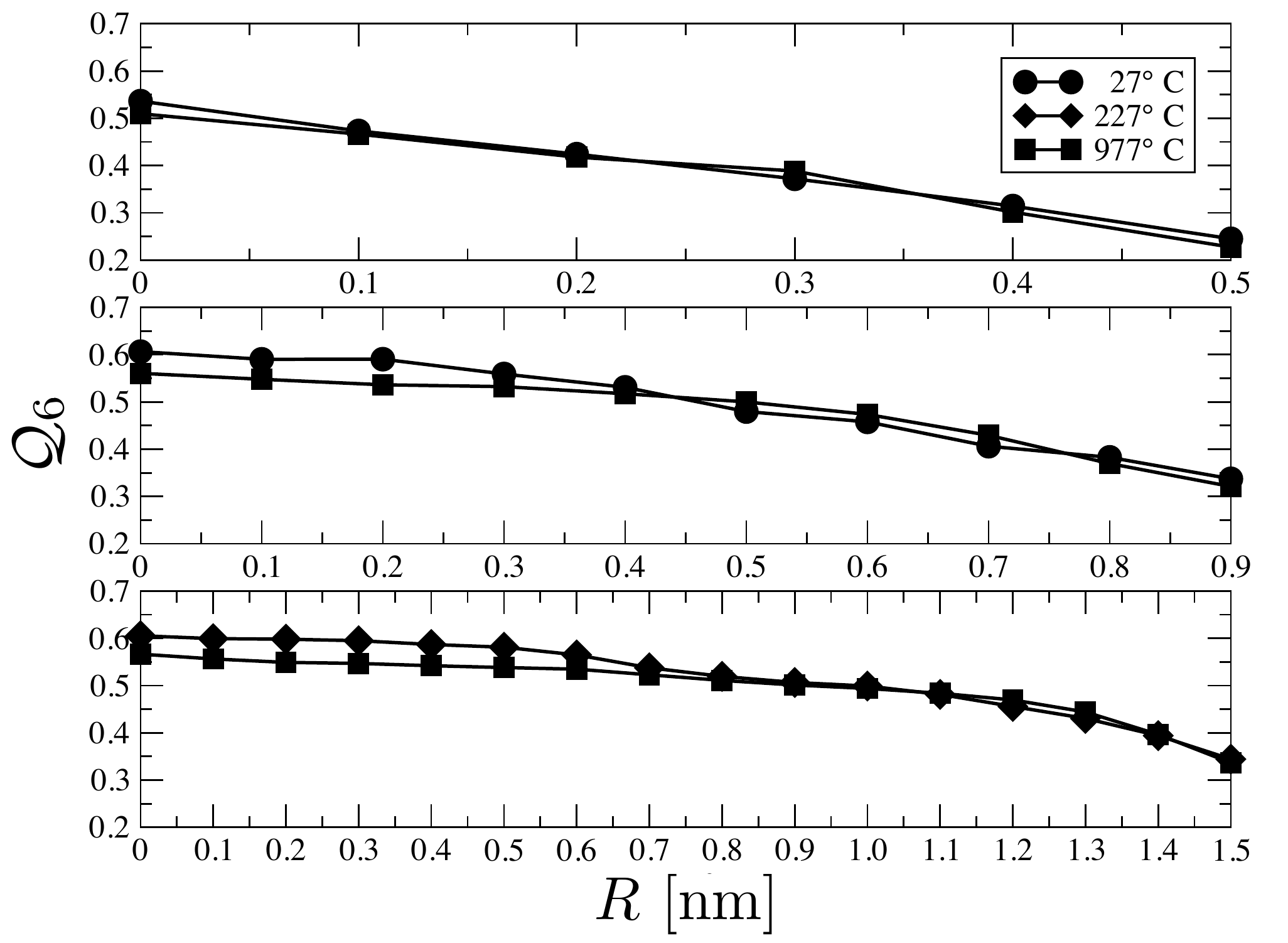}
 \end{center}
 \caption{${\mathcal Q}_6(R)$  for the $0.8$~nm (top),  $1.3$~nm (middle) and $1.8$~nm (bottom) nanoparticles. For each nanoparticle, the ${\mathcal Q}_6(R)$ is computed at the same $T$ as in Fig.~\ref{fig:snapshots}}\label{fig:Q6R}
 \end{figure}

\textcolor{black}{We further investigated the structure of the crystalline metastable states by computing the shell-by-shell ${\mathcal Q}_6$. This quantity, denoted by the symbol ${\mathcal Q}_6(R)$, is obtained by limiting the sums in Eq.~\ref{eq:AppQlmNano} to the atoms laying in the spherical layer of thickness $0.1$~nm at the distance $R$ from the centre of the nanoparticle. In Fig.~\ref{fig:Q6R} is shown the ${\mathcal Q}_6(R)$ for the three nanoparticles at the same $T$ as in Fig.~\ref{fig:snapshots}. As a first remark, Fig.~\ref{fig:Q6R} confirms the observation that the degree of order decreases in going from the center to the periphery of the nanoparticle. This is consistent with the results of previous works. \cite{PhysRevLett.93.226104} However, Fig.~\ref{fig:Q6R} also indicates that there is a qualitative difference in the structure of nanoparticles of different size.}
\textcolor{black}{ For the nanoparticle of size $0.8$~nm we notice that the ${\mathcal Q}_6(R)$ decreases monotonically with $R$. The trend is very similar both at low and high $T$. On the contrary, already for the nanoparticle of size $1.3$~nm, we observe that the ${\mathcal Q}_6(R)$ is characterized by one plateau region in the domain $0$~nm~$\leq R \leq 0.5$~nm. Beyond this region, the ${\mathcal Q}_6$ quickly decreases, till reaching the value of $\sim 0.3$ at the Si/a-SiO$_2$ interface. For the largest nanoparticles, at low $T$ we observe two plateau regions, one in the domain $0$~nm~$\leq R \leq 0.5$~nm\ and another, at lower ${\mathcal Q}_6$, in the domain $0.5$~nm~$< R \leq 1.1$~nm. At higher $T$ we have only one plateau region in the domain $0$~nm$\leq R \leq 1.1$~nm. In both cases, beyond $R = 1.1$~nm the ${\mathcal Q}_6$ goes very quickly to the value of $\sim 0.35$. These observations indicate that there is a qualitative difference between the structure of the crystalline nanoparticle with their size, namely that there is a threshold below which the ordered nanoparticle does not have a crystalline core with a homogeneous degree of order. This difference is preserved also at higher $T$}

%

\textcolor{black}{Once having identified the nature of the low and high ${\mathcal Q}_6$ metastable states, and having analyzed their structure, we turn to the analysis of the order-disorder phase transition with the size of the nanoparticle and the temperature of the system.} 
Our simulations (see Fig.~\ref{FOfQ6}) provide the following qualitative sharp picture: for small nano-particles (${\mathcal R}^* = 0.8-1.3$~nm) at low temperature (T $<  727 ^\circ$C) the most stable configuration corresponds to the amorphous phase, while the crystalline state is found to be more stable at higher temperatures. On the contrary, for larger particles (${\mathcal R}^* \geq 1.8$~nm) this behavior is inverted, resulting similar to bulk-like conditions: at low temperatures (T $< 977 ^\circ$~C) the crystalline phase is the most stable one, while the amorphous phase is preferred at higher temperatures. Interestingly enough, for small nano-particles the equilibrium temperature (i.e. the temperature at which the free energy of the disordered and ordered phase are the same) decreases with the increase of the size of the nano-particle. This is indeed an effect of the steady increase of stability of the crystalline phase with respect to the disordered one with the size of the nano-particle.
The results of our simulations provide the following picture, consistent with the experimental results: \cite{jap-95_3723, jap-103_114303, Wakayama1998124, jap-83_2228, Wang2006486, jp-19_225003} at low annealing temperature the nano-particles are small and amorphous as, due to the inversion of stability with respect to bulk-like systems, this is thermodynamically the most stable phase; \textcolor{black}{at moderately higher temperatures the average size and the size distribution of the nano-particles are slightly changed (the average size is slightly increased). The largest particles in the sample transform from amorphous to crystalline, the most stable phase for large nanoparticles  at this $T$.} By further increasing the temperature the average size of the nano-particles increases significantly and the larger ones tend toward the crystalline state (i.e. they follow the change in stability from disorder to order, as induced by their growing size). On the other hand, the smaller particles undergo a amorphous-to-crystalline transition due to the increase of the temperature and the inversion of the stability with respect to the bulk-like system. Even in this case they eventually crystallize. The model above is based on the experimental observation of the dependency of the average size and size distribution of the nanoparticles on the temperature. \cite{jp-19_225003} \textcolor{black}{Our results bring to the conclusion that the observed crystallization of the nanoparticles with the increase of the temperature is due to the interplay of the effect of the temperature on their size and the inversion of the relative stability of the amorphous and crystalline phase with the temperature for small nanoparticles.}

\section{Conslusions}
\label{Sec:Conclusions}

In this paper we investigated the relative stability of the amorphous vs crystalline nanoparticles of size ranging between $0.8$ and $1.8$~nm. We found that, at variance from bulk systems, at low $T$ small nanoparticles are amorphous and they undergo to an amorphous-to-crystalline phase transition at high $T$. On the contrary, large nanoparticles recover the bulk-like behavior: crystalline at low $T$ and amorphous at high $T$. 

\textcolor{black}{We also investigated the structure of the crystalline nanoparticle. Our results, in agreement with previous works, \cite{PhysRevLett.93.226104} demonstrate that this kind of nanoparticle are formed by an ordered core surrounded by a disordered periphery. However, they also indicate that the details of the structure of the crystalline core depend on the size of the nanoparticle}

\acknowledgements
The authors wish to acknowledge the SFI/HEA Irish Centre for High-End Computing (ICHEC) for the provision of computational facilities. One of the authors (S. M.) acknowledges SFI Grant 08-IN.1-I1869 and the European Community under the Marie Curie Intra-European Fellowship for Career Development grant number 255406 for the financial support. One of the authors (S.O.) acknowledges SimBioMa for financial support.

\newpage
%
%

%
%

\end{document}